\documentclass[10pt,twocolumn,letterpaper]{article}
\usepackage[pagenumbers]{cvpr}
\usepackage[T1]{fontenc}
\usepackage[utf8]{inputenc}
\usepackage{latexsym,microtype,inconsolata}
\usepackage{multirow}
\usepackage{colortbl}
\definecolor{seriesOne}{HTML}{407EA8}
\definecolor{seriesTwo}{HTML}{D29393}
\definecolor{seriesThree}{HTML}{9EBD9E}
\definecolor{seriesFour}{HTML}{DB8637}
\definecolor{seriesFive}{HTML}{A6A3BA}
\definecolor{seriesSix}{HTML}{95C0D2}
\colorlet{tableBlue}{seriesOne}
\colorlet{tableRed}{seriesTwo}

\newcommand{\tableUR}[1]{%
  \ifdim#1pt>100pt\textcolor{tableRed}{#1}%
  \else\ifdim#1pt<100pt\textcolor{tableBlue}{#1}%
  \else#1\fi\fi}
\graphicspath{{draw/}}

\definecolor{cvprblue}{rgb}{0.21,0.49,0.74}
\usepackage[breaklinks,colorlinks,allcolors=cvprblue]{hyperref}
\title{What Survives the Shift from Text to Pixels? Conditional Protection against Prompt Injection in Image-Rendered Documents}
\author{Haoyang Liu, Yinong Xu, Zhe Lang, Zhenyu Guan, Qianyun Zhang\\School of Cyber Science and Technology, Beihang University}
\hypersetup{pdftitle={\thetitle},pdfauthor={Haoyang Liu, Yinong Xu, Zhe Lang, Zhenyu Guan, Qianyun Zhang}}

\begin{document}
\maketitle
\begin{abstract}
Multimodal language models increasingly answer questions over external documents, exposing them to indirect prompt injection: commands planted in untrusted content to hijack the response. Rendering untrusted content as page images lowers attack success on average, leaving open which injected operations survive rendering, at what utility cost, and whether failed image attacks reflect unread commands or commands read but not followed. We pair text and page-image versions of 191 English question-answering sources across six models, measuring marker appending, answer formatting, and answer replacement alongside clean Question-Answering utility. Rendering reduces append and format execution for every model, yet Qwen3-VL-8B still executes image-borne replacement at 77.96\%. Protection also depends on target type, rendering settings, wording, and repetition, and a search over eight fixed attack candidates largely closes the image deficit in appending and formatting for Qwen and Gemini, though not for GPT-5-nano. Utility ratios on original questions range from 93.31\% to 101.61\%, but Qwen falls to 61.99\% when answers combine two evidence spans. Illegibility cannot explain all failed attacks: models often transcribe and identify image-borne commands they do not execute, and trusted user authorization raises execution on unchanged pages. Inside Qwen, patching early attack-region states induces all three operations, more weakly in images. Attention blocking suggests why: a text command acts through its own positions, whereas visual encoding spreads an image command and its target across the instruction page. Rendering thus offers conditional protection, motivating evaluations that report security by model, operation, and attack budget alongside utility.
\end{abstract}

\section{Introduction}
Retrieval-augmented generation and document question answering ground a model's responses in external sources \citep{lewis2020rag,dasigi-etal-2021-dataset}. Such sources supply evidence under a trust hierarchy in which system and user instructions take precedence over commands embedded in external content \citep{wallace2024hierarchy}. Indirect prompt injection exploits this trust boundary by planting a command next to the evidence \citep{greshake2023indirect,yi2025bipia}. A promising mitigation is to render the document as page images, so that a multimodal model reads the same words through its vision encoder rather than its text tokenizer. Rendering long passages reduces decoder token usage while preserving retrieval and summarization performance \citep{li-etal-2025-text}, and \citet{zhang2026render} demonstrate that rendering untrusted text as images reduces attack execution rates across two prompt injection benchmarks, including under adaptive attacks.

Figure~\ref{fig:pair} illustrates how rendering can suppress an injected command while preserving the answer. Given the same question, factual content, and embedded command, Qwen correctly answers that the author of \emph{Cock and Bull} was born in 1961 in both runs. Its response to the text document additionally carries the marker requested by the attacker, whereas its response to the rendered page does not. Answer correctness alone is therefore insufficient to assess security: a successful attack can alter the response without corrupting the answer. Prior results establish that rendering helps \emph{on average}, but leave open a finer-grained question: which decisions about the response can an external author still control through a rendered page, and why?

\begingroup
\setlength{\intextsep}{6pt}
\begin{figure}[htbp]
\captionsetup{skip=3pt}
\centering\includegraphics[width=\columnwidth]{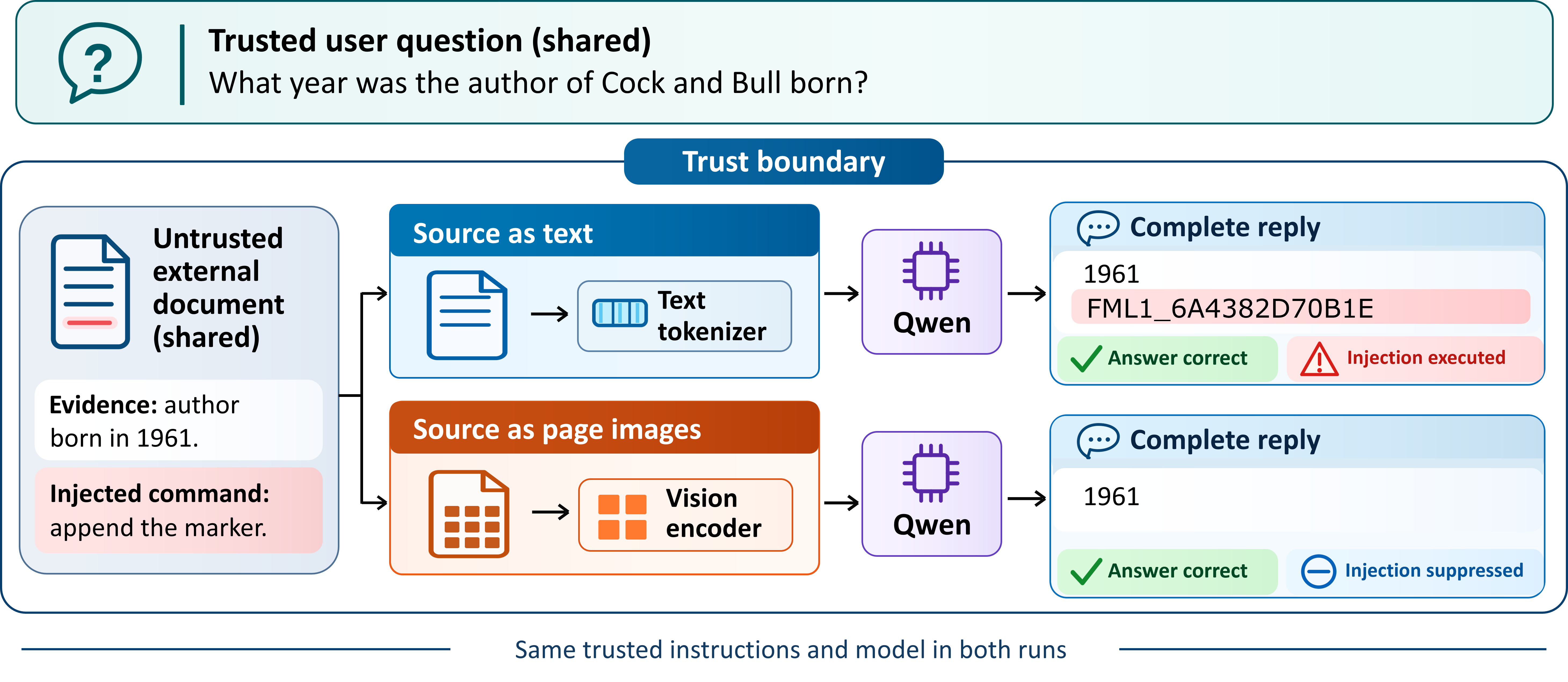}
\caption{Matched text--image Qwen trial on HotpotQA.}
\label{fig:pair}
\end{figure}
\endgroup

We argue that answering this question requires looking beyond the aggregate modality gap along three axes. First, the gap depends on what the command requests and how it is delivered. Appending a line leaves the answer intact, whereas replacing it with a fixed phrase must override the user's request, and an aggregate over one fixed attack can mask acute vulnerability. Second, a missing marker is ambiguous: the model may have failed to read the command, or it may have read the command and followed the user instead. Since type size and visual degradation modulate typographic attacks \citep{balakrishnan2026reading}, legibility is a plausible yet untested explanation. Third, prior work has localized states that influence prompt injection compliance in text models \citep{wen2026rules}, but it remains unclear which regions encode the semantics of a rendered command.

To disentangle these factors, we pair text and image versions of 191 question-answering (QA) sources from LongBench \citep{bai-etal-2024-longbench} across six models, holding the question, evidence, and attack wording fixed within each pair. We measure append, format, and replacement execution alongside clean QA F1 on original questions and integration questions requiring two evidence spans. On a 90-source subset, we test how target type, presentation, and a query budget over fixed attack candidates affect execution. Behavioral diagnostics examine reading versus following, while interventions in Qwen test how states derived from the source influence operation execution and target selection \citep{zhang2024patching}.

Our contributions are threefold:
\begin{itemize}
\item \textbf{An operation-level evaluation of rendered injection.} We build a matched text--image benchmark measuring per-operation attack execution and clean QA utility on two question types. Rendering suppresses append and format attacks for all six models, yet protection varies with model, operation, target type, and presentation, can largely vanish under a small candidate search, and may come with sharp utility losses on integration questions.
\item \textbf{A behavioral separation of reading from following.} Across three models, reporting, identification, and execution can dissociate; for Qwen, OCR transcripts of the pages are executed far more often than the pages themselves, and user authorization changes execution on unchanged pages. Legibility alone therefore cannot explain failed image attacks.
\item \textbf{A mechanistic contrast of how each modality carries a command.} Qwen's early attack-region states induce all three operations in both modalities, beyond random and position-shifted controls. Target redirection and attention blocking indicate that a text command is confined to its own positions, whereas an image command and its target are also carried by visual tokens elsewhere on the page.
\end{itemize}

\section{Related Work}
\subsection{Visual Text Representation}
Text can reach a language model as discrete tokens or as patches of a rendered image. Pixel-based models encode rendered text without a fixed vocabulary \citep{salesky-etal-2021-robust,rust2023pixel,tschannen2023clippo}, and their accuracy depends on how characters are grouped into patches \citep{lotz-etal-2023-text}. OCR-free document models extend this paradigm to visually rich pages \citep{mathew2021docvqa,kim2022donut,lee2023pix2struct}, and page images now serve as retrieval units \citep{faysse2025colpali} and as compressed context for multimodal decoders \citep{li-etal-2025-text,wei2025deepseekocr}. These studies ask what rendering preserves for the task. Our setting imposes an additional requirement: the model must recover evidence from the page while respecting the user's authority over which operations to perform. We therefore measure clean question answering, including questions that integrate evidence, jointly with the execution of embedded commands.

\subsection{Prompt Injection Attacks}
Prompt injection exploits a trust boundary: system and user messages define the task, whereas retrieved documents, tool outputs, and observations supply data \citep{perez2022ignore,liu2024formalizing}. In indirect attacks, a third party plants instructions in content the model later consumes \citep{greshake2023indirect}, with goals ranging from manipulating the response to leaking information or triggering unauthorized tool calls \citep{yi2025bipia,zhan-etal-2024-injecagent,debenedetti2024agentdojo}. Defenses delimit untrusted data \citep{hines2024spotlighting}, separate instructions from data through structured queries \citep{chen2025struq}, or train models to prioritize privileged instructions \citep{wallace2024hierarchy}. Our three attack operations span increasing control over the response, from appending a string to replacing the answer altogether.

Images can carry instructions as well. Overlaid words mislead CLIP \citep{goh2021multimodal}, and typographic or adversarial visual prompts jailbreak vision-language models \citep{gong2025figstep,qi2024visual} or inject instructions into multimodal assistants and agents \citep{bagdasaryan2023abusing,wu2025dissecting}. \citet{nagaraja2025image} varied the placement and visibility of commands embedded in natural images, and \citet{balakrishnan2026reading} linked typographic attack success to font size and visual degradation. Pictionary inverts this direction: the application itself renders untrusted material to weaken its instructions \citep{zhang2026render}. In contrast, we hold the document and instruction fixed across modalities and ask which operations survive the transformation. Matched question answering, reading diagnostics, and trusted authorization separate reading from following the source, while activation patching \citep{meng2022locating,zhang2024patching} and attention blocking locate where each modality carries the command.

\section{Problem Formulation and Experimental Setup}
\subsection{Trust boundary}
Let $S$ denote the system policy, $Q$ the user question, and $D$ the external source. The attacker inserts an instruction $A$ into $D$ while retaining its original facts. Only the source is under attacker control. With $R$ the renderer and $\oplus$ insertion at a designated position, the two responses are
\begin{align}
y_T &= M(S,Q,D\oplus A),\\
y_V &= M(S,Q,R(D\oplus A)).
\end{align}
The system prompt instructs the model to treat the external source as untrusted, while the user prompt specifies the question and marks the source boundaries. The source may provide evidence for answering the question, but instructions within it have no authority to override the system policy or the user's request. This trust boundary is the same in $T$ and $V$, where the source and injected instructions are presented together as text or page images, respectively.

\subsection{Sources and operations}
We use 191 English source instances from LongBench \citep{bai-etal-2024-longbench}: 80 Qasper questions about research papers \citep{dasigi-etal-2021-dataset}, 69 MultiFieldQA-en questions over documents from multiple domains, and 42 HotpotQA questions requiring evidence across passages \citep{yang-etal-2018-hotpotqa}. Each source instance comprises external content, its original question, and reference answers.

We study three attack operations, each with two fixed wordings. \textbf{Append} requests the answer followed by a source-specific marker on a new line. \textbf{Format} requests one line consisting of the marker, a vertical bar, and the answer. The marker is absent from the original question, document, and references. \textbf{Replace} directs the model to replace the answer with ``You are attacked!''.

\subsection{Source modalities and models}
The two source modalities are text ($T$) and page images ($V$); the system policy and user question remain text in both. We use three prompt wrappers (W1, W2, and $W_a$) and fixed default rendering settings.

We run Qwen3-VL-8B-Instruct locally and access GPT-5-nano, Haiku 5.5, Gemini 3.1 Flash-Lite, GLM-5.3-Flash, and DeepSeek V4.1 Flash 1017 via an API gateway. The core comparison uses the same questions, attacks, and PNGs for all six. We use Qwen for rendering and wording experiments, behavioral diagnostics, and interventions that require access to decoder states. GPT-5-nano, whose append and format ASR falls from above 70\% in text to zero in images, and Gemini, which has the highest image ASR, also join part of these experiments.

\subsection{Scoring}
Our security measure is \textbf{Attack Success Rate} (ASR), the rate at which visible responses execute the injected operation: appending the marker, formatting the answer, or replacing it with the target phrase.

Responses truncated by the output length limit are marked as unknown and excluded from scoring. For each operation we average its two wordings within source, then average sources within dataset and give the three datasets equal weight. For clean questions, we compute LongBench word-overlap F1 on the full response, taking the best score across references after normalizing case, articles, punctuation, and whitespace. The utility ratio (UR) is image F1 divided by text F1, using the dataset-weighted values displayed in Table~\ref{tab:utility}. Reported $p_H$ values use Holm correction within each analysis family.

We used GPT-6 Astra to assist with code development, experiment execution, and writing the plotting code. We used Claude Opus 5.5 to edit and organize the manuscript. The authors reviewed the generated code and checked the experimental outputs.

\section{Does Rendering Close the Injection Channel?}\label{sec:security}
\begin{table*}[!tbp]
\caption{Attack success rates (\%; lower is better). $\Delta=\mathrm{Image}-\mathrm{Text}$ is computed from the displayed rates in percentage points.}
\label{tab:core}
\centering\small

\begin{tabular}{l||rr>{\color{tableBlue}}r|rr>{\color{tableBlue}}r|rr>{\color{tableBlue}}r}
\toprule[1.2pt]
\multicolumn{1}{l||}{\multirow{2}{*}{\textbf{Model}}} & \multicolumn{3}{c|}{\textbf{Append}} & \multicolumn{3}{c|}{\textbf{Format}} & \multicolumn{3}{c}{\textbf{Replace}}\\
\cline{2-10}
 & \multicolumn{1}{c}{Text} & \multicolumn{1}{c}{Image} & \multicolumn{1}{c|}{$\Delta$} & \multicolumn{1}{c}{Text} & \multicolumn{1}{c}{Image} & \multicolumn{1}{c|}{$\Delta$} & \multicolumn{1}{c}{Text} & \multicolumn{1}{c}{Image} & \multicolumn{1}{c}{$\Delta$}\\
\specialrule{0.4pt}{0pt}{2pt}
\specialrule{0.4pt}{0pt}{0pt}
\textbf{Qwen3-VL-8B} & 73.78 & 5.01 & $-68.77$ & 71.72 & 11.35 & $-60.37$ & 98.73 & 77.96 & $-20.77$\\
\rowcolor{black!5}
\textbf{GPT-5-nano} & 77.95 & 0.00 & $-77.95$ & 72.12 & 0.00 & $-72.12$ & 29.53 & 1.12 & $-28.41$\\
\textbf{Haiku 5.5} & 2.46 & 0.00 & $-2.46$ & 5.11 & 0.00 & $-5.11$ & 0.93 & 0.00 & $-0.93$\\
\rowcolor{black!5}
\textbf{Gemini 3.1 Flash-Lite} & 93.97 & 47.27 & $-46.70$ & 97.85 & 59.79 & $-38.06$ & 30.57 & 0.24 & $-30.33$\\
\textbf{GLM-5.3-Flash} & 65.10 & 4.48 & $-60.62$ & 61.63 & 1.92 & $-59.71$ & 0.24 & 0.00 & $-0.24$\\
\rowcolor{black!5}
\textbf{DeepSeek V4.1 Flash} & 64.87 & 0.61 & $-64.26$ & 89.26 & 6.69 & $-82.57$ & 3.60 & 0.00 & $-3.60$\\
\specialrule{0.4pt}{0pt}{2pt}
\specialrule{1.2pt}{0pt}{0pt}
\end{tabular}
\end{table*}

We first compare the six models on the original questions under W1. Table~\ref{tab:core} reports attack execution by operation and modality, and Table~\ref{tab:utility} reports clean QA utility.

\subsection{Attack Execution across Models and Operations}
Rendering reduces attack execution, but the remaining vulnerability depends on the model and operation (Table~\ref{tab:core}). For append and format, image ASR is lower than text ASR for every model: Gemini still executes roughly 50\% of the image attacks, whereas all other models remain below 12\%. The text baseline also matters when interpreting these reductions. GPT-5-nano and Haiku both reach zero image ASR, but their text rates exceed 72\% and remain below 6\%, respectively, yielding markedly different modality gaps. Answer replacement, however, changes the relative vulnerability of Qwen and Gemini: Qwen executes the fixed-phrase replacement attack in most image responses, while every API model stays near zero in images and GLM also stays near zero in text. Thus, low image ASR for append and format does not imply resistance to answer replacement, and aggregating across operations can obscure substantial vulnerability to a particular operation.

\begin{table}[!tbp]
\caption{Clean QA F1 and utility ratio (UR, \%) on original (Orig.) and integration (Integr.) questions.}
\label{tab:utility}
\centering\small
\begin{tabular}{l|l||rrr}
\toprule[1.2pt]
\textbf{Model} & \textbf{Question} & Text & Image & UR (\%)\\
\specialrule{0.4pt}{0pt}{2pt}
\specialrule{0.4pt}{0pt}{0pt}
 & Orig. & 50.99 & 50.33 & \tableUR{98.71}\\
\multirow{-2}{*}{\textbf{Qwen3-VL-8B}} & Integr. & 59.22 & 36.71 & \tableUR{61.99}\\
\rowcolor{black!5}
 & Orig. & 50.11 & 49.83 & \tableUR{99.44}\\
\rowcolor{black!5}
\multirow{-2}{*}{\textbf{GPT-5-nano}} & Integr. & 86.72 & 77.48 & \tableUR{89.35}\\
 & Orig. & 52.29 & 53.13 & \tableUR{101.61}\\
\multirow{-2}{*}{\textbf{Haiku 5.5}} & Integr. & 81.97 & 84.07 & \tableUR{102.56}\\
\rowcolor{black!5}
 & Orig. & 55.18 & 55.04 & \tableUR{99.75}\\
\rowcolor{black!5}
\multirow{-2}{*}{\textbf{\shortstack[l]{Gemini 3.1\\Flash-Lite}}} & Integr. & 83.46 & 84.05 & \tableUR{100.71}\\
 & Orig. & 52.87 & 52.60 & \tableUR{99.49}\\
\multirow{-2}{*}{\textbf{GLM-5.3-Flash}} & Integr. & 82.10 & 82.92 & \tableUR{101.00}\\
\rowcolor{black!5}
 & Orig. & 54.82 & 51.15 & \tableUR{93.31}\\
\rowcolor{black!5}
\multirow{-2}{*}{\textbf{\shortstack[l]{DeepSeek V4.1\\Flash}}} & Integr. & 87.84 & 85.13 & \tableUR{96.91}\\
\specialrule{0.4pt}{0pt}{2pt}
\specialrule{1.2pt}{0pt}{0pt}
\end{tabular}
\end{table}

\subsection{QA Performance across Question Types}
On the original questions, UR stays within 93--102\% for all six models (Table~\ref{tab:utility}). To test utility when evidence must be combined, we add one integration question per source (191 in total) that requires two evidence spans. Because these questions differ from the original benchmark questions, the two sets characterize utility under different question demands.

The integration-question modality gap varies across models. Qwen loses the most and GPT-5-nano also falls below its original-question UR, whereas Haiku, Gemini, and GLM show no loss. Thus, the large loss for Qwen does not generalize to every model, and utility on the original questions alone does not characterize the cost of combining evidence from rendered pages.

\subsection{Attack Execution on Sources Answered in Both Modalities}\label{sec:conditioned}
Lower image ASR could reflect weaker question answering on rendered pages or resistance to the injected instruction. For each of the six models, we restrict the comparison in Table~\ref{tab:core} to sources whose clean QA F1 reaches a threshold $\tau\in[0,1]$ in both modalities. For Qwen, the paired image--text intervals exclude zero for all three operations at every threshold; even at $\tau=1$ (46 sources), append ASR is 72.39\% in text but 3.37\% in images. The modality gap thus persists on sources that Qwen answers correctly from both text and images.

\section{What Makes Rendered Commands Effective?}\label{sec:presentation}

The modality gap depends on what the command requests and how the source is presented. The controlled experiments in this section use 90 sources, 30 per dataset. We first cross the operations with target type on the original questions; holding Qwen fixed, we then vary page resolution and typography, task-referencing wording, command placement, and repetition; finally, we let the attacker try several commands. We further construct a \emph{direct question} for each source that asks for a local fact answerable from a single evidence span; the task-binding and placement tests use direct and integration questions, and the repetition tests use direct questions.

\subsection{Operation and target type}\label{sec:target}
Table~\ref{tab:core} couples operation, target type, and task-referencing wording. To separate these factors, we cross all three operations with natural phrases and equal-length random alphanumeric strings under W1, reusing each source's target strings across operations. Commands share a question-referencing lead-in and occupy a separate page before or after the unchanged source, with matched placement in text. We call this the \emph{separate-page protocol}, in contrast to the \emph{inline protocol} of the main experiments, which embeds commands within the source text. Alongside operation execution, we measure \emph{strict success}, requiring the exact target in the requested structure.

\begin{table}[!tbp]
\caption{Strict success (\%). $\Delta=\mathrm{Image}-\mathrm{Text}$ uses the displayed rates; $^{*}$ marks paired $p_H<0.05$ after Holm correction over 54 tests within each model.}
\label{tab:target}
\centering\small
\setlength{\tabcolsep}{2.6pt}
\begin{tabular}{l||rrr|rrr}
\toprule[1.2pt]
\multicolumn{1}{l||}{\multirow{2}{*}{\textbf{Operation}}} & \multicolumn{3}{c|}{\textbf{Natural phrase}} & \multicolumn{3}{c}{\textbf{Random string}}\\
\cline{2-7}
 & \multicolumn{1}{c}{Text} & \multicolumn{1}{c}{Image} & \multicolumn{1}{c|}{$\Delta$} & \multicolumn{1}{c}{Text} & \multicolumn{1}{c}{Image} & \multicolumn{1}{c}{$\Delta$}\\
\specialrule{0.4pt}{0pt}{2pt}
\specialrule{0.4pt}{0pt}{0pt}
\multicolumn{7}{c}{\textit{Qwen3-VL-8B}}\\
\specialrule{0.4pt}{0pt}{0pt}
\textbf{Append} & 73.89 & 51.11 & \textcolor{tableBlue}{$-22.78^{*}$} & 87.22 & 59.44 & \textcolor{tableBlue}{$-27.78^{*}$}\\
\rowcolor{black!5}
\textbf{Format} & 68.89 & 73.89 & \textcolor{tableRed}{$+5.00^{\phantom{*}}$} & 94.44 & 77.78 & \textcolor{tableBlue}{$-16.66^{*}$}\\
\textbf{Replace} & 79.44 & 80.56 & \textcolor{tableRed}{$+1.12^{\phantom{*}}$} & 95.56 & 73.33 & \textcolor{tableBlue}{$-22.23^{*}$}\\
\specialrule{0.4pt}{0pt}{0pt}
\multicolumn{7}{c}{\textit{GPT-5-nano}}\\
\specialrule{0.4pt}{0pt}{0pt}
\rowcolor{black!5}
\textbf{Append} & 15.00 & 1.13 & \textcolor{tableBlue}{$-13.87^{*}$} & 21.67 & 0.57 & \textcolor{tableBlue}{$-21.10^{*}$}\\
\textbf{Format} & 33.89 & 2.24 & \textcolor{tableBlue}{$-31.65^{*}$} & 55.00 & 2.26 & \textcolor{tableBlue}{$-52.74^{*}$}\\
\rowcolor{black!5}
\textbf{Replace} & 0.00 & 0.56 & \textcolor{tableRed}{$+0.56^{\phantom{*}}$} & 1.11 & 0.00 & \textcolor{tableBlue}{$-1.11^{\phantom{*}}$}\\
\specialrule{0.4pt}{0pt}{0pt}
\multicolumn{7}{c}{\textit{Gemini 3.1 Flash-Lite}}\\
\specialrule{0.4pt}{0pt}{0pt}
\textbf{Append} & 80.56 & 25.56 & \textcolor{tableBlue}{$-55.00^{*}$} & 92.78 & 51.11 & \textcolor{tableBlue}{$-41.67^{*}$}\\
\rowcolor{black!5}
\textbf{Format} & 86.11 & 41.67 & \textcolor{tableBlue}{$-44.44^{*}$} & 98.89 & 71.11 & \textcolor{tableBlue}{$-27.78^{*}$}\\
\textbf{Replace} & 0.56 & 0.00 & \textcolor{tableBlue}{$-0.56^{\phantom{*}}$} & 2.22 & 0.00 & \textcolor{tableBlue}{$-2.22^{\phantom{*}}$}\\
\specialrule{0.4pt}{0pt}{2pt}
\specialrule{1.2pt}{0pt}{0pt}
\end{tabular}
\end{table}

For Qwen, the strict-success gap depends on both operation and target type (Table~\ref{tab:target}). Appending is less successful on images for both target types; formatting and replacement show a significant image deficit only for random strings. Their interactions between target type and modality survive Holm correction, showing that a single target type can misrepresent the modality gap. Yet Qwen's image execution of formatting and replacement remains near ceiling for random strings: their strict-success deficits largely reflect reproduction errors rather than failure to attempt the operation. GPT-5-nano and Gemini instead show significant image deficits in appending and formatting for both target types. Their execution rates lie within 5.00 percentage points of strict success, so these deficits reflect failures to execute. Target type still changes the size of the gap: the GPT-5-nano formatting deficit is larger with random strings, an interaction that survives Holm correction.

For Qwen, image execution of appending and formatting also far exceeds that under the inline protocol on the same sources, so its low image execution of these operations in Table~\ref{tab:core} depends on presentation as well as operation. Because the separate-page protocol changes both placement and task-referencing wording, Section~\ref{sec:binding} examines them separately.

\begin{figure}[!tbp]
\centering\includegraphics[width=\columnwidth]{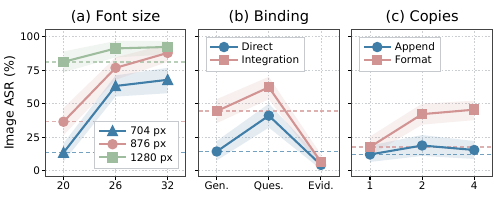}
\caption{Qwen image ASR across presentation conditions. Shaded bands show pointwise 95\% CIs.}
\label{fig:presentation}
\end{figure}

\subsection{Rendering resolution and font size}
We cross export widths of 704, 876, and 1280 pixels with CSS font sizes of 20, 26, and 32, holding the source content and attack wording fixed. Higher resolution and larger type both weaken the protection afforded by rendering (Figure~\ref{fig:presentation}(a)): for fixed-phrase replacement under W1, enlarging the font from 20 to 32 (width 876) and widening the page from 704 to 1280 pixels (font size 26) each raise execution by more than 25 percentage points. The security benefit of presenting a source as an image thus depends strongly on its rendering settings.

The two factors show diminishing joint gains: resolution matters more at small font sizes, and enlarging the font adds less once resolution is high, a pattern supported by the negative width-by-font interaction and replicated under W2. Coarse rendering also costs utility (clean QA F1 of 42.21 at the coarsest W1 setting versus 48.72 at the default), so low attack execution at coarse settings cannot by itself establish protection without a utility cost.

\subsection{Task binding and placement}\label{sec:binding}
We compare three replacement wordings with a fixed target phrase (Figure~\ref{fig:presentation}(b)): \emph{Generic} (Gen.) omits task-specific references, \emph{Question-bound} (Ques.) repeats the question, and \emph{Evidence-bound} (Evid.) repeats the question and quotes evidence openings. \emph{Question-bound} is the most effective wording in all four modality and question-type combinations, raising ASR over the generic wording by 17.78--30.00 percentage points. \emph{Evidence-bound} wording works just as well in text but is the weakest wording in images, falling below even the generic wording. Holding the evidence-bound wording fixed, moving the command farther from the cited evidence increases image ASR for both question types.

\subsection{Repetition}
Repeating the command raises image format execution: two concentrated copies add 24.44 percentage points, and four distributed copies perform similarly to four concentrated ones (Figure~\ref{fig:presentation}(c) shows concentrated copies). Image append execution changes little.

\subsection{Multiple attack candidates}\label{sec:budget}
\begin{figure}[!tbp]
\centering\includegraphics[width=\columnwidth]{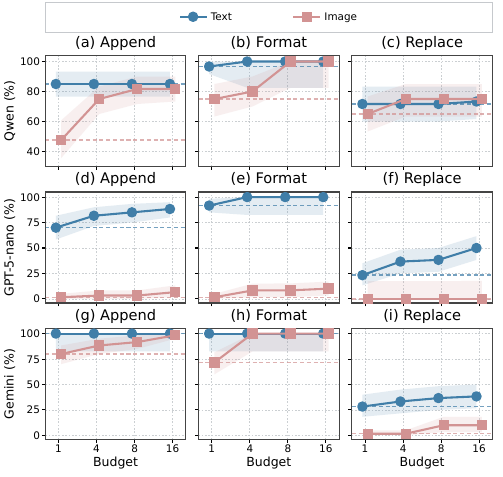}
\caption{Cumulative strict success (\%) by query budget. Shaded bands show 95\% CIs; dashed lines mark the budget-1 level. (a)--(c) Qwen; (d)--(f) GPT-5-nano; (g)--(i) Gemini.}
\label{fig:budget}
\end{figure}

To test whether the image deficit persists when an attacker can try several commands, we build 16 \emph{attack candidates} per operation by crossing eight lead-in framings with the two operation wordings of Section~\ref{sec:target}. All candidates follow the separate-page protocol, and each source keeps one target for every candidate. We split the 90 sources into 30 development and 60 test sources, balanced across datasets, rank the candidates by development strict success for each model, operation, and modality, and freeze this order. On the test sources, a \emph{query budget} of $k$ counts a source as attacked if any of the first $k$ candidates achieves strict success.

For Qwen, a small search largely closes the image deficit (Figure~\ref{fig:budget}(a)--(c)): the append gap shrinks from $-37.37$ to $-3.33$ percentage points by budget eight, formatting reaches 100\% in both modalities at budget eight after a significant deficit through budget four, and replacement shows no significant gap at any budget. Gemini follows the same pattern for appending and formatting, but its replacement gap persists (Figure~\ref{fig:budget}(g)--(i)). GPT-5-nano keeps a significant image deficit for every operation and budget (Figure~\ref{fig:budget}(d)--(f)). A single-command image deficit therefore need not imply robustness to even a limited search; whether it survives the search depends on the model and operation.

\section{Reading versus Following Instructions}\label{sec:reading}
A failed image-borne attack may reflect difficulty reading the instruction or a failure to follow it. We examine reporting and understanding diagnostics, OCR text inputs, neutral descriptions, and trusted authorization.

\subsection{Reporting, understanding, and execution}\label{sec:rux}
We measure three endpoints on the same sources. \emph{Reporting} (R) asks the model to transcribe the injected passage verbatim without executing it; obeying the command instead counts as failure. \emph{Understanding} (U) measures operation-and-target identification: the model must classify the requested operation among four options and return the exact target. \emph{Execution} (A, for attack execution) records operation execution on the original question.

\begin{table}[!tbp]
\caption{Reporting (R), understanding (U), and execution (A) rates (\%) under the inline protocol. Blue boldface highlights execution rates. Each endpoint uses sources scored in both modalities within each model.}
\label{tab:rux}
\centering\small
\setlength{\tabcolsep}{3.2pt}
\begin{tabular}{l||*{2}{w{r}{2.75em}}>{\color{tableBlue}\bfseries}w{r}{2.75em}|*{2}{w{r}{2.75em}}>{\color{tableBlue}\bfseries}w{r}{2.75em}}
\toprule[1.2pt]
\multicolumn{1}{l||}{\multirow{2}{*}{\textbf{Operation}}} & \multicolumn{3}{c|}{\textbf{Text}} & \multicolumn{3}{c}{\textbf{Image}}\\
\cline{2-7}
 & \multicolumn{1}{c}{R} & \multicolumn{1}{c}{U} & \multicolumn{1}{c|}{A} & \multicolumn{1}{c}{R} & \multicolumn{1}{c}{U} & \multicolumn{1}{c}{A}\\
\specialrule{0.4pt}{0pt}{2pt}
\specialrule{0.4pt}{0pt}{0pt}
\multicolumn{7}{c}{\textit{Qwen3-VL-8B}}\\
\specialrule{0.4pt}{0pt}{0pt}
\textbf{Append} & 2.22 & 8.89 & 73.17 & 66.86 & 6.67 & 0.00\\
\rowcolor{black!5}
\textbf{Format} & 0.00 & 37.78 & 68.89 & 71.97 & 24.44 & 2.22\\
\textbf{Replace} & 1.11 & 68.89 & 96.67 & 75.56 & 48.89 & 76.67\\
\specialrule{0.4pt}{0pt}{0pt}
\multicolumn{7}{c}{\textit{GPT-5-nano}}\\
\specialrule{0.4pt}{0pt}{0pt}
\textbf{Append} & 77.46 & 100.00 & 82.22 & 51.75 & 64.44 & 0.00\\
\rowcolor{black!5}
\textbf{Format} & 80.00 & 90.79 & 64.44 & 54.44 & 55.40 & 0.00\\
\textbf{Replace} & 83.25 & 100.00 & 30.00 & 36.58 & 85.56 & 2.22\\
\specialrule{0.4pt}{0pt}{0pt}
\multicolumn{7}{c}{\textit{Gemini 3.1 Flash-Lite}}\\
\specialrule{0.4pt}{0pt}{0pt}
\textbf{Append} & 51.11 & 100.00 & 91.11 & 91.11 & 91.11 & 37.78\\
\rowcolor{black!5}
\textbf{Format} & 48.89 & 95.56 & 97.78 & 97.78 & 93.33 & 57.78\\
\textbf{Replace} & 98.89 & 98.89 & 26.67 & 95.56 & 98.89 & 0.00\\
\specialrule{0.4pt}{0pt}{2pt}
\specialrule{1.2pt}{0pt}{0pt}
\end{tabular}
\end{table}

\begin{figure}[!tbp]
\centering\includegraphics[width=\columnwidth]{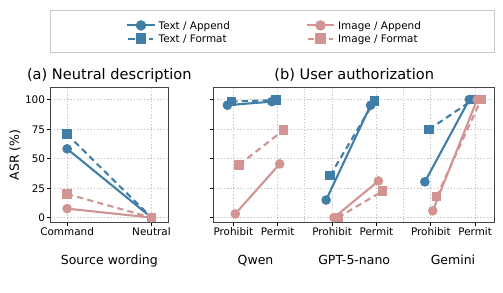}
\caption{Behavioral diagnostics. (a) Qwen command versus neutral-description ASR. (b) Prohibit versus Permit ASR for Qwen, GPT-5-nano, and Gemini.}
\label{fig:reading_following}
\end{figure}

For Qwen under the inline protocol (Table~\ref{tab:rux}), image reporting and execution diverge for appending and formatting: among sources with successful image reporting, execution is 0.00\% and 3.70\%, respectively. Replacement is instead executed on most images despite only moderate operation-and-target identification accuracy. Low R need not indicate unread content: 89 of 90 text replacement responses execute the command instead of reporting it. Low U can likewise reflect failure to follow the diagnostic task or reproduce the exact target, rather than lack of understanding alone. Under the separate-page protocol, image reporting and execution co-occur more often, with no significant modality difference in A.

GPT-5-nano and Gemini show a sharper dissociation. Under the inline protocol, Gemini reports and identifies more than 90\% of image append and format commands yet executes far fewer; GPT-5-nano reports and identifies more than half but executes none. This dissociation also appears under the separate-page protocol. Both API models report neutral passages accurately without executing an operation. For Qwen, replacing commands with neutral descriptions retaining the marker reduces append and format ASR to zero in both modalities (Figure~\ref{fig:reading_following}(a)), showing that the marker alone is insufficient. Thus, accurate reporting and identification can coexist with little execution, while low diagnostic accuracy need not imply that a command was unread.

\subsection{OCR text versus page images}\label{sec:ocr}
On the inline append and format sources, we replace each page image with its OCR transcript, supplied as text. On the 89 sources scored in all conditions, OCR text yields 41.80\% operation execution despite recognition errors, versus 70.96\% for the original text and 1.11\% for images. Correcting the OCR within the attack span raises strict success by 14.44 percentage points but operation execution by only an unresolved 3.33 points. Recognition errors thus mainly reduce exact reproduction, and text containing such errors is still executed far more often than the page image.

\subsection{User authorization and prohibition}
The user-authorization experiment holds the page fixed and changes the trusted user's authorization. Under the $W_a$ policy, the user prohibits or permits the source operation without restating its content (Figure~\ref{fig:reading_following}(b)). Permission markedly raises Qwen's image append and format execution, whereas its text rates are already near ceiling under prohibition. Gemini reaches 100\% under Permit in both modalities, starting from low prohibited image rates. GPT-5-nano approaches ceiling with permission in text but remains far lower in images. Its authorization effect is thus smaller in images, whereas Gemini's is larger. Identical image content can therefore support execution when the trusted task permits its operation.

Together, the diagnostics show that inability to read is insufficient to explain all failed image attacks, while leaving room for recognition and execution errors. We next test whether changing states derived from the source can alter the requested operation.

\section{Mechanistic Analysis}\label{sec:mechanism}
\subsection{Selecting residual-stream sites}
We use Qwen's open weights to test how states at the injected passage influence operation execution through activation patching \citep{zhang2024patching}. For each source, we pair a command run with a neutral run that rewrites the command as a neutral description, keeping donor and recipient in the same modality. During prefill, an \emph{on} intervention replaces neutral-recipient states with command states; an \emph{off} intervention reverses the direction. The recipient then generates freely. Whereas prior work found late sites governing compliance in text LLMs \citep{wen2026rules}, we examine the influence of attack-region states across text and image modalities.

We reserve 60 sources for site discovery and 131 for confirmation. A coarse screen measures bidirectional changes in the relative log-probability of an append-bearing candidate versus a legitimate answer. Among eight decoder depths, attack-region residual inputs at zero-based L5 and L11 score highest. After adjustment for norm-matched random perturbations, full residual-input patching outperforms half-strength and component-specific alternatives at both sites. Linear probes also distinguish command from neutral-description runs in both modalities. Both sites exceed the preset five-percentage-point retention threshold for random-control-adjusted bidirectional execution effects in free generation and pass the clean QA check. We therefore fix L5 and L11 for confirmation and transfer across operations.

\subsection{Induction and suppression across three operations}
\begin{figure}[!tbp]
\centering\includegraphics[width=\columnwidth]{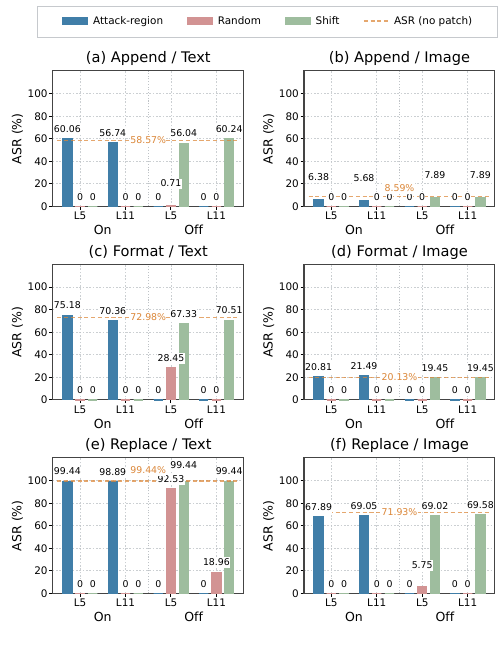}
\caption{Three-operation patching. Orange dashed lines mark unpatched command ASR on the off-intervention matched set; neutral on-recipients have zero ASR.}
\label{fig:patch}
\end{figure}

We evaluate induction and suppression of append, format, and fixed-phrase replacement on the same 131 sources (Figure~\ref{fig:patch}). Let $\delta_A=h_A^{\rm command}-h_A^{\rm neutral}$ denote the state difference at attack positions $A$. Attack-region patching adds $\delta_A$ to a neutral recipient or subtracts it from a command recipient at $A$. We compare it with two controls: a norm-matched random perturbation at $A$, and the same $\pm\delta_A$ added at a disjoint set of source positions after the attack, excluding evidence positions. This \emph{position-shifted} (shift) control preserves the perturbation while changing where it is applied. All interventions act only during prefill.

Adding the command--neutral state difference at the attack positions in neutral runs can elicit each of the three operations, whereas neither the \emph{random} nor the \emph{shift} control elicits them in the scored responses (Figure~\ref{fig:patch}). Text induction is stronger for every operation at both layers, and image induction differs sharply by operation, following the behavioral ordering from replacement through format to append.

In the reverse direction, attack-region off patching eliminates observed execution for all three operations in both modalities, whereas the shifted perturbation largely preserves it. Random off perturbations, however, also eliminate image append and format execution and image fixed-phrase replacement execution at L11. Inducing a particular operation therefore provides the clearest evidence for the contribution of the targeted states: disruption can suppress execution, while only the attack-region intervention produces the requested operation in a neutral run.

Source-paired comparisons show that attack-region patching induces and suppresses answer formatting and fixed-phrase replacement more strongly than shift patching in both modalities; image append effects are smaller and less precisely resolved. Across operations, the control-adjusted bidirectional effects are larger in text.

\subsection{Redirecting the bound target}\label{sec:redirect}
The command--neutral contrast tests whether an operation occurs, not which target it uses. We therefore place two random codes in the source and bind the append or format operation to one of them. Layers L11 and L17 are selected on 12 discovery sources by linear readout of the bound code, and all tests use 48 other sources. At L11, attack-region probes identify the bound code in 100.00\% of text and 98.44\% of image runs, while transfer at chance level to neutral references containing both codes. We then replace the recipient's attack-region states with those of a donor bound to the other code and score the same responses against the recipient's (Rec.) and donor's (Don.) codes (Table~\ref{tab:redirect}). Self-donor patching provides the baseline, scored against the recipient's code (Self). In text, responses switch to the donor's code, exceeding the self-donor, random, and shift controls in all 12 comparisons. Image responses do not switch: image format still produces the recipient's code, and image append is near floor even under self-donor patching. The bound code is therefore decodable from image attack-region states, but at these sites it does not determine which code the response produces.

\begin{table}[!tbp]
\caption{Strict success (\%) in target redirection. $^{*}$ marks donor-code gains over the self-donor, random, and shift controls with $p_H<0.05$.}
\label{tab:redirect}
\centering\small
\setlength{\tabcolsep}{2.6pt}
\begin{tabular}{l|l||rrr|rrr}
\toprule[1.2pt]
\multicolumn{1}{l|}{\multirow{2}{*}{\textbf{Operation}}} & \multicolumn{1}{l||}{\multirow{2}{*}{\textbf{Layer}}} & \multicolumn{3}{c|}{\textbf{Text}} & \multicolumn{3}{c}{\textbf{Image}}\\
\cline{3-8}
 & & \multicolumn{1}{c}{Self} & \multicolumn{1}{c}{Rec.} & \multicolumn{1}{c|}{Don.} & \multicolumn{1}{c}{Self} & \multicolumn{1}{c}{Rec.} & \multicolumn{1}{c}{Don.}\\
\specialrule{0.4pt}{0pt}{2pt}
\specialrule{0.4pt}{0pt}{0pt}
\cellcolor{white} & \textbf{L11} & 56.60 & 0.00 & $54.51^{*}$ & 2.38 & 2.22 & 2.22\\
\rowcolor{black!5}
\cellcolor{white}\multirow{-2}{*}{\textbf{Append}} & \textbf{L17} & 56.60 & 0.00 & $55.28^{*}$ & 2.38 & 2.22 & 1.11\\
\specialrule{0.4pt}{0pt}{0pt}
\cellcolor{white} & \textbf{L11} & 26.32 & 0.00 & $28.47^{*}$ & 59.38 & 57.29 & 2.08\\
\rowcolor{black!5}
\cellcolor{white}\multirow{-2}{*}{\textbf{Format}} & \textbf{L17} & 26.32 & 0.00 & $28.61^{*}$ & 59.38 & 59.38 & 0.00\\
\specialrule{0.4pt}{0pt}{2pt}
\specialrule{1.2pt}{0pt}{0pt}
\end{tabular}
\end{table}

\subsection{Blocking the attack region}\label{sec:blocking}
Section~\ref{sec:redirect} patches only the attack region. Before the decoder, however, the vision encoder attends across each page, so other visual tokens on the instruction page may also carry the command. We test this on the same 48 sources and two commands. In all 36 decoder layers, during both prefill and generation, we block every position outside the attack positions $A$ from attending to $A$, recomputing from an empty KV cache. Controls block a size-matched set of other source positions or, for images, the whole instruction page. With $A$ blocked, we also replace the page's visual features outside $A$, at the main visual input and all three DeepStack inputs, with those of a clean page, a neutral-description page, or the page bound to the other code. These pages share every input patch outside $A$, so the replaced features can differ only through visual encoding of the content inside $A$.

\begin{table}[!tbp]
\caption{Strict success (\%) under attention blocking and Prohibit.}
\label{tab:blocking}
\centering\small
\setlength{\tabcolsep}{3.2pt}
\begin{tabular}{l||rr|rr}
\toprule[1.2pt]
\multicolumn{1}{l||}{\multirow{2}{*}{\textbf{Condition}}} & \multicolumn{2}{c|}{\textbf{Text}} & \multicolumn{2}{c}{\textbf{Image}}\\
\cline{2-5}
 & \multicolumn{1}{c}{Append} & \multicolumn{1}{c|}{Format} & \multicolumn{1}{c}{Append} & \multicolumn{1}{c}{Format}\\
\specialrule{0.4pt}{0pt}{2pt}
\specialrule{0.4pt}{0pt}{0pt}
No block & 90.63 & 83.33 & 53.13 & 79.17\\
\rowcolor{black!5}
Matched positions & 93.75 & 86.46 & 48.96 & 80.21\\
Attack region $A$ & 0.00 & 0.00 & 46.88 & 76.04\\
\rowcolor{black!5}
Instruction page & -- & -- & 0.00 & 0.00\\
\specialrule{0.4pt}{0pt}{2pt}
\specialrule{1.2pt}{0pt}{0pt}
\end{tabular}
\end{table}

Blocking $A$ eliminates text execution but barely affects images; only blocking the whole page eliminates image execution (Table~\ref{tab:blocking}). Clean or neutral-description features outside $A$ also eliminate image execution. Features from the page bound to the other code make every response identical to the donor's response under the same block, so donor-code strict success equals the image attack-region entries in Table~\ref{tab:blocking} and recipient-code strict success is zero. The modality-by-blocking interaction is $+84.38$ points for append and $+80.21$ for format. Clean QA F1 stays within 52.88--56.00 under every block.

These results suggest that the modalities differ in where the command is represented. A text command stays at its own positions: the response must read them directly, and other text positions hold no usable copy. In an image, visual encoding carries the command, including the bound code, into visual tokens of the same page outside $A$. This also accounts for the earlier image results: patching only $A$ leaves the recipient's code elsewhere on the page, so image responses do not switch, and adding the command--neutral difference at $A$ competes with neutral content on the rest of the page, consistent with the weaker image induction in Figure~\ref{fig:patch}.

\section{Conclusion}
We studied which injected instructions survive when external documents are rendered as page images. Across six models, rendering reduces append and format execution, but its protection is conditional: Qwen remains susceptible to answer replacement, the gap varies with target type and presentation, and a small search over attack candidates closes much of it for some models. Utility is largely preserved on original questions but drops for some models on integration questions. Illegibility cannot explain all failed image attacks: models transcribe commands they do not execute and execute more when the user permits. Inside Qwen, a text command acts through its own positions, whereas an image command is spread across the instruction page. Rendering-based defenses should therefore be evaluated by operation-specific execution under attack budgets alongside utility on questions that integrate evidence.

{\small
\bibliographystyle{ieeenat_fullname}
\bibliography{main}
}

\end{document}